\documentclass{IEEEmce}

\usepackage[colorlinks,urlcolor=blue,linkcolor=blue,citecolor=blue]{hyperref}

\usepackage{upmath}
\usepackage{epstopdf}
\usepackage{color}

\jvol{XX}
\jnum{XX}
\paper{8}
\jmonth{May/June}
\publisheddate{00 xxxx 0000}
\currentdate{00 xxxx 0000}
\jname{IEEE Consumer Electronics Magazine}
\pubyear{2022}
\doiinfo{MCE.2022.Doi Number}

\begin{document}

\sptitle{Special Section: Security and Privacy-Aware Emerging Computing}
\editor{Editor: Name, xxxx@email}

\title{AI-Oriented Two-Phase Multi-Factor Authentication in SAGINs: Prospects and Challenges}

\author{Bin Yang}
\affil{Zhejiang Lab, Hangzhou, China}

\author{Shanyun Liu}
\affil{Zhejiang Lab, Hangzhou, China}

\author{Tao Xu}
\affil{Zhejiang Lab, Hangzhou, China}

\author{Chuyu Li}
\affil{Zhejiang Lab, Hangzhou, China}

\author{Yongdong Zhu}
\affil{Zhejiang Lab, Hangzhou, China}

\author{Zipeng Li}
\affil{Nanchang University, Nanchang, China}

\author{Zhifeng Zhao}
\affil{Zhejiang Lab, Hangzhou, China}

\markboth{Security and Privacy-Aware Emerging Computing}{AI-Oriented Two-Phase Multi-Factor Authentication in SAGINs: Prospects and Challenges}

\begin{abstract}
Space-air-ground integrated networks (SAGINs), which have emerged
as an expansion of terrestrial networks, provide flexible access,
ubiquitous coverage, high-capacity backhaul, and emergency/disaster
recovery for mobile users (MUs). While the massive benefits brought by SAGIN may improve the quality of service, unauthorized access to SAGIN entities is potentially dangerous. At present, conventional crypto-based authentication
is facing challenges, such as the inability to provide continuous and
transparent protection for MUs. In this article, we propose an AI-oriented two-phase multi-factor
authentication scheme (ATMAS) by introducing intelligence
to authentication. The satellite and network control center collaborate
on continuous authentication, while unique spatial-temporal features, including service features and geographic features, are utilized to enhance the system security. Our further security analysis and performance evaluations show that ATMAS has proper security characteristics which can meet various security requirements. Moreover, we shed light on lightweight and efficient authentication mechanism design through a proper combination of spatial-temporal factors.
\end{abstract}

\maketitle

\enlargethispage{10pt}

\chapterinitial{With the prosperity} of the fifth-generation (5G) mobile communication
networks, the number of 5G subscribers has already reached an enormous
scale, leading to a significant shift in people's daily lives.
Nowadays, academia and industry are turning their attention to emerging
next-generation systems. It is envisioned that the sixth-generation (6G) communications will support five application scenarios: Enhanced Mobile Broadband Plus, Big Communications, Secure Ultra-Reliable Low-Latency Communications, Three-Dimensional Integrated Communications, and Unconventional Data Communications \cite{Ye20Space}. Space-air-ground integrated networks (SAGINs) leverage the advantages of satellite networks (including geosynchronous Earth orbit (GEO) satellites, medium Earth orbit (MEO) satellites, low Earth orbit (LEO) satellites, and their mutual links), high altitude platforms, and terrestrial communication systems. SAGINs are therefore considered as a promising architecture to support ubiquitous, seamless, reliable and high-data-rate services anytime and anywhere in 6G communication networks. Remarkably, practical experiments
and ambitious projects have been initiated to offer ubiquitous Internet
services through SAGINs, such as Google Loon, Thales Stratobus, OneWeb, and
SpaceX.

SAGIN is a highly heterogeneous and multidimensional network compared to conventional terrestrial or satellite networks. It has wide coverage, broadcast channels, and various network entities, including networking infrastructure and connected Internet of Thing (IoT) devices. However, its wide coverage feature brings trust and safety issues. Specifically, the wireless signal in SAGIN propagates mainly in free space. In other words, not only authorized users could receive the information but
also malicious users could obtain wireless power and retrieve secure
information from power leakage in wireless signals. Besides, intrinsic
trust and data reliability issues may arise in SAGIN during multi-hop
transmissions among distrustful entities within each segment or across
segments. Due to the lack of unified security precautions, interconnected
intelligent devices can be vulnerable to various cyber attacks, and
traceable data provenance is difficult \cite{Wang22Blockchain}.
Therefore, merits including security, privacy, and intelligence have
become the crux of determining whether SAGIN can continue to evolve healthily.

As the first security perimeter,  authentication is a pivotal method
to identify the legitimacy of IoT devices that access the network,
thereby enabling real-time monitoring and promoting collaborative sharing
\cite{Yao20Toward,Kumar22Light}. Authentication can be classified into one-shot authentication and continuous authentication based on the duration of the authentication process. One-shot authentication is an elementary
authentication mechanism that identifies IoT devices using crypto-based
techniques such as passcodes, PINs, and fingerprints. However, it only protects the security of SAGIN during the initial
access process and cannot guarantee security during the operational phase.
With the proliferation of intelligent IoT devices, especially wearable
devices like fitness bands and augmented reality (AR) glasses, personal
behavioral and physiological biometrics can be easily collected by
built-in sensors, which promotes continuous authentication from concept
to implementation. As an enhancement and supplement to one-shot authentication,
continuous authentication defenses against attacks in the background
without user intervention by constantly validating access devices
according to the historical features of users. To cope with the new features
of SAGINs, it becomes increasingly essential to combine one-shot authentication
for quick access and continuous authentication for sustainable security
together, supporting convenient access and security robustness of
various kinds of communication entities \cite{Liu18Space}.

To adequately utilize the collected data from IoT devices, adopting multi-factor authentication (MFA) is promising. This means that multiple heterogeneous validation methods can be combined intelligently to
grant or deny access reliably \cite{Ometov19Challenges}. In MFA,
three types of factor groups, i.e., knowledge factor, ownership factor
and biometric factor, are available to connect an individual with the
established credentials. Knowledge factor, ownership factor and biometric
factor refer to factors like passwords, tokens and behavioral patterns,
respectively. Specifically, spatial-temporal features, such as geographical
position, Doppler shift and traffic volume, are likely to be utilized
in the continuous authentication phase due to the critical roles of
SAGINs in providing varieties of vertical applications by connecting
enormous heterogeneous devices, machines, and industrial processes.
Through MFA by artificial intelligence (AI) techniques, including supervised
learning, unsupervised learning, and reinforcement learning, trusted
communications and services can be accomplished to adapt to dynamic environments.

Driven by the limitations of conventional authentication mechanisms
and the induced demands in an integrated network, this article presents
a novel authentication framework for SAGINs, where an AI-oriented
two-phase multi-factor authentication scheme (ATMAS) is proposed.
In ATMAS, a machine-learning-based continuous authentication which
refers to 'Phase II' is performed to intelligently grant or deny access
reliably by capturing user-profiles and traits, following a conventional
cryptographic authentication named 'Phase I'. The contributions of this article are summarized
as follows:
\begin{itemize}
  \item We propose the ATMAS based on the characteristics of SAGINs. For the ATMAS, four communication entities, i.e., mobile user (MU), base station (BS), satellite, and network control center (NCC) are considered, in which satellite and NCC cooperate with each other to implement authentication.
  \item Through the design of ATMAS, we leverage the unique spatial-temporal features of SAGIN in Phase II, including service and geographic features.  This allows us to ensure the security of the SAGIN from login to logout.
  \item To evaluate the robustness and validate the security of the proposed scheme, we conduct security analysis by analyzing the security characteristics. Moreover, from the perspective of performance evaluation, we shed light on lightweight and efficient authentication mechanism design by a proper combination of spatial-temporal factors.
\end{itemize}


The remainder of this article is organized as follows. Challenges
of authentication for the SAGINs are first introduced. Then we present
the advantages of AI-enhanced multi-factor continuous authentication.
The design of the proposed ATMAS and its security analysis are described in the following Sections. In addition, simulation results are presented to evaluate the performance of authentication accuracy. Finally,
concluding remarks and future research directions are provided.

\begin{figure*}
\centerline{\includegraphics[width=35pc]{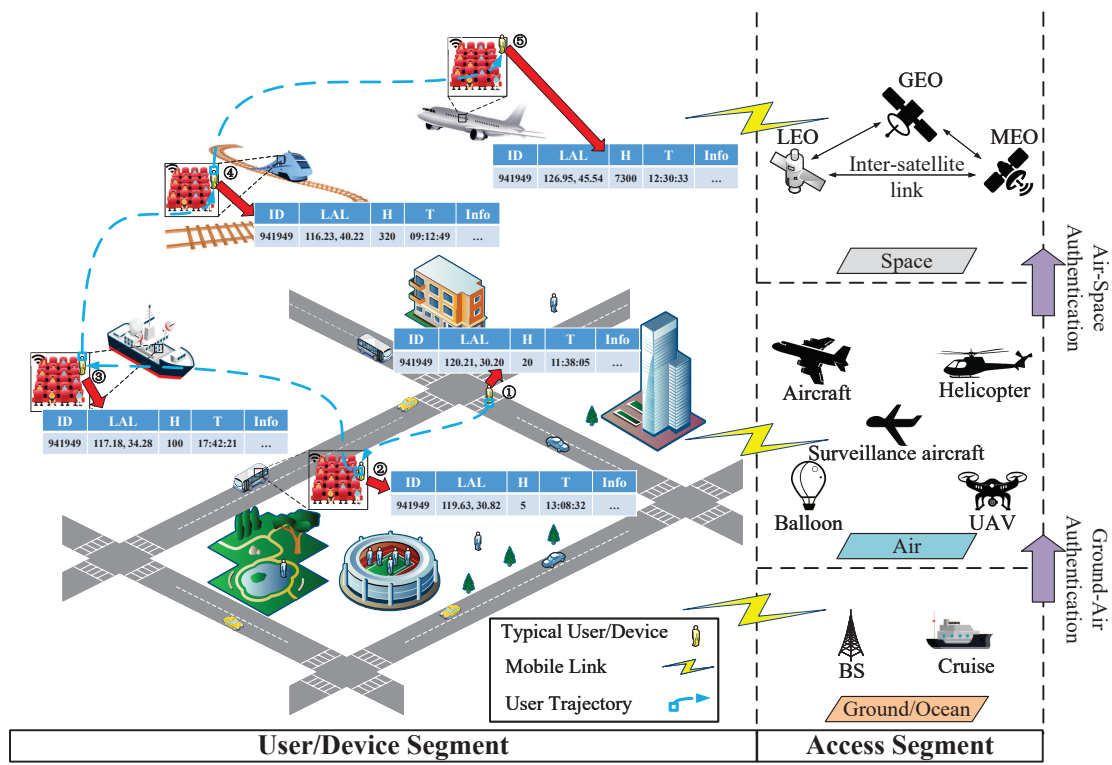}}
\caption{\label{fig:Architecture}Network Architecture of the SAGIN. The user/device segment shows the trajectory of a MU and the changes of its location, traffic and biometric. The ground/air/space access segment illustrates all kinds of access entities, including BSs, high-altitude platforms and satellites.}
\end{figure*}

\section{Challenges of Authentication for the SAGIN\label{sec:2 Motivations}}

Comprised of inter-satellite and inter-satellite links, the SAGIN
is a highly open wireless system. As satellites are exposed to the
open environment, it is easy to cause attacks such as link hijacking
and entity counterfeiting, which seriously affects the secure communication
of the SAGIN. Reliable access control plays a vital role in distinguishing
the source nodes and addressing identity-based attacks such as spoofing
and Sybil attacks. In general, access control can be decomposed into
user authentication and authorization. Authentication is a process
that verifies that someone or something is who they say they are and
authorization is the security process that determines a user or service's
level of access. The latter can be accomplished
by methods like enforcing allow or deny rules based on the user's
authorization level, e.g., general user, super user and administrator,
which is relatively less complicated in the implementation stage. However, authentication brings
about a number of SAGIN-specific research questions that should be
carefully considered in this article.

\textbf{Long Propagation and Processing Latencies.} In SAGIN, the introduction of high-altitude satellites leads to unidirectional propagation delays of around 15 ms, even for the links between LEO satellites
and IoT devices. This results in round-trip delays
of approximately 50 ms for IoT devices, which significantly reduces quality of service (QoS)
in latency-sensitive scenarios. Additionally, due to high mobility,
the propagation delay is not permanent and varies with the satellite's
position. Apart from propagation delay, conventional cryptographic
authentication requires increased communication and computation overhead
to cope with the increasing requirements of security, leading to
long processing latencies \cite{Fang20Fast}. Such prolonged latencies are intolerable for scenarios like disaster rescue
and military missions. Therefore, lightweight and selectable authentication mechanisms
are urgently needed for these latency-sensitive applications.

\textbf{Insufficiency of Authentication Uniformity.} As mentioned before,
the SAGIN is a highly heterogeneous network supported by
different communication protocols. Each network entity in the SAGIN
encompasses a tremendous number of IoT devices with various interfaces
for control and management, which requires a unified and pragmatic
authentication mechanism when IoT devices access the SAGIN
from different locations. While a unified authentication
framework has been proposed for both the Third Generation Partnership Project
(3GPP) and non-3GPP access in \cite{3GPP,Cui21Edge}, sensitive data-related
authentication must be enforced to be implemented in the core network,
which mitigates efficiency. As a result, a new unified authentication
scheme is necessary to adapt to the heterogeneity of the
SAGIN.

\textbf{Inability to Achieve Real-time and Transparent Authentication.}
The traditional crypto-based authentication verifies the legitimacy
of IoT devices only at the beginning of login based on a password,
a personal identification number (PIN), or a secret pattern. However,
these methods are susceptible to guessing, video capture, and spoofing
as users often choose a simple password for convenience. Moreover,
traditional one-shot authentication mechanisms cannot verify
whether the logged-on user is the initially authenticated one due
to a lack of physiological and behavioral biometrics. To overcome
these shortcomings, users have to identify themselves by re-entering
the password periodically, which declines the experience and satisfaction
of services. In short, traditional authentication mechanisms lead
to either security vulnerabilities or inconvenience for the SAGIN.
To achieve privacy protection in the background, new mechanisms should
be designed for real-time and transparent security provisioning.

To put it succinctly, it is imperative to enhance the current authentication mechanism for the SAGIN-specific system.

\begin{figure}
\centerline{\includegraphics[width=17.5pc]{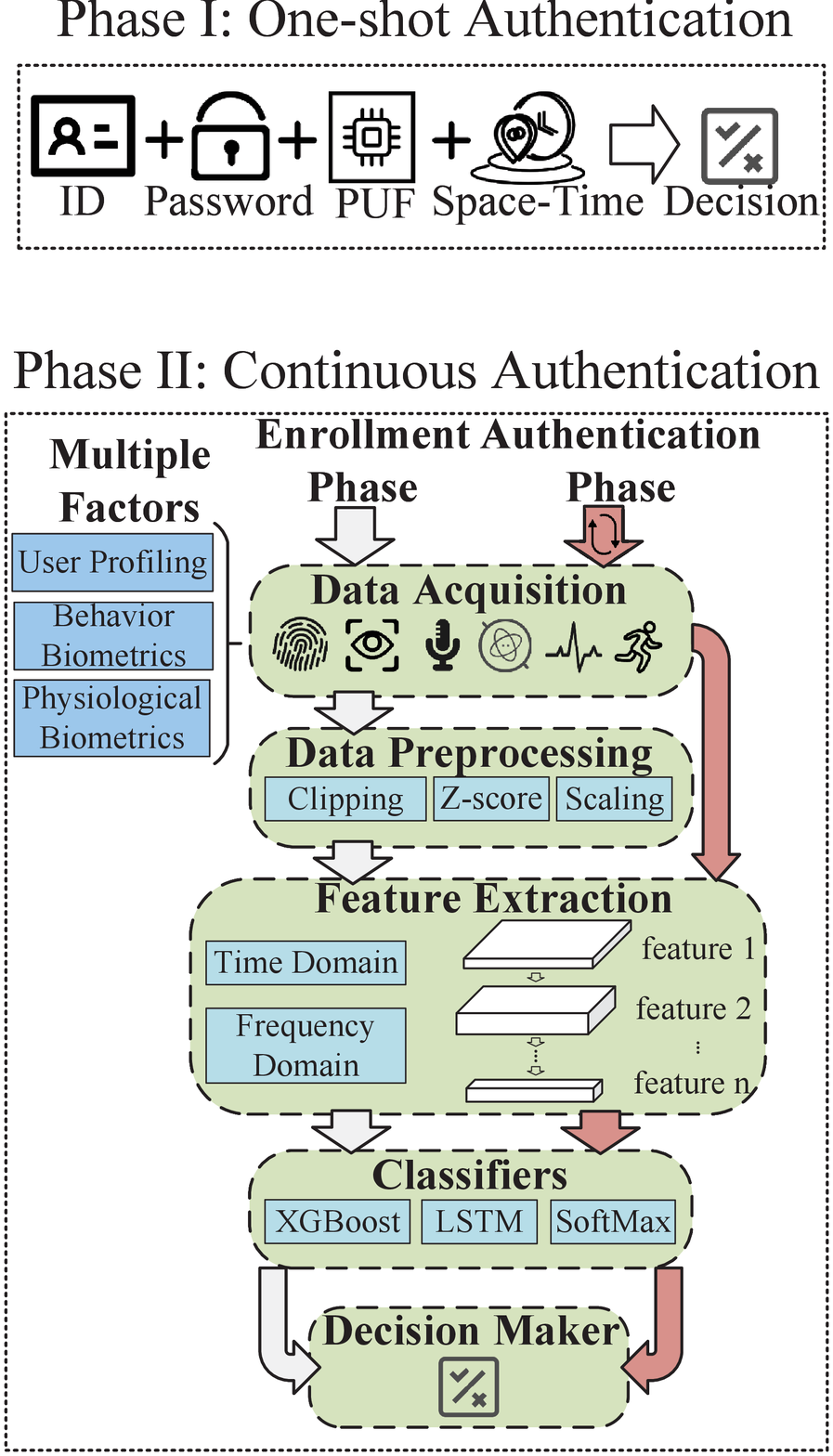}}
\caption{\label{fig:Phase}Authentication Phase of the ATMAS. In Phase I, a crypto-based authentication is performed. In Phase II, an AI-based continuous authentication including enrollment phase and authentication phase is proposed.}
\end{figure}

\section{AI-Enhanced Multi-Factor Continuous Authentication }

In the era of IoT, the SAGIN system has become more complex than traditional information and communication technology platforms. To ensure the SAGIN security from login to logout during IoT device activity, a continuous authentication mechanism is indispensable to
fulfill security requirements without user intervention. Typically,
continuous authentication validates legitimate users based on behavioral
and physiological biometrics using built-in sensors on IoT devices. In
recent years, due to advancements in storage and computational resources,
and the increasing diversity of sensors, continuous authentication has become
more effective and accurate by collecting, storing, and analyzing
massive amounts of data \cite{Abuhamad21Sensor}.

The current authentication mechanisms can be divided into two
categories: single-factor authentication and MFA, based on the features or authentication factors applied. Single-factor
authentication is a standard, low-security method of authentication
that requires matching only one factor, such as password, to a username
to get access to a system. However, single-factor authentication provides
a limited level of security because it is susceptible to guessing,
video capture, and spoofing. MFA is an enhanced authentication
mechanism that jointly conducts authentication utilizing multiple factors. Factors in authentication can be classified into five
categories: \textbf{i.} something a user knows, \textbf{ii.}
something a user possesses, \textbf{iii.} something a user is, \textbf{iv.}
something a user does and \textbf{v.} somewhere a user is \cite{Modarres22An}.
With proper factor selection and combination strategies, an efficient
and high-security authentication mechanism can be achieved.

By extracting the unique features of users from the abundant collected
data, AI algorithms including machine learning and deep learning are
leveraged to enhance the security of the SAGIN system. As illustrated
in Fig. \ref{fig:Phase}, the pipeline of AI-based authentication
mechanisms mainly consists of data acquisition, data preprocessing,
feature extraction, classification and decision. Data acquisition
involves sensors that sample real-world parameters and convert them to electrical signals, which are then converted to digital values. Data
preprocessing is a critical procedure that distill high-quality data
from raw data, which is generally incomplete, noisy, inconsistent,
and redundant. To reduce noise and align output
data, methods like clipping, Z-score, and scaling are necessary \cite{Liang20Behavioral}.
Feature extraction means extracting user features that present
one's identity or behavior from the collected dataset. For classification,
the goal is to learn a mapping function that predicts label information
for a given behavior sequence with minimal biases. Finally, through
evaluations of a verification system, the ultimate decision can be
made to grant access to legitimate users and deny access to impostors.
In this article, we envision ATMAS to address challenges for the SAGIN. The advantages of ATMAS are
summarized as follows.

\textbf{Continuous Security.} The proposed ATMAS not only authenticates at the initial stage of login but also provides seamless protection to legitimate devices in the background by adopting AI-enhanced algorithms.

\textbf{High Flexibility.} We utilize a conventional cryptographic
authentication named `Phase I' to quickly grant or deny service
requester access, and an AI-aided continuous authentication named `Phase II'
to intelligently verify legitimacy during service operation.
The SAGIN system operator is flexible in choosing whether to use only `Phase I' or
to select a suitable number of authentication factors based on security
requirements.

\textbf{High Robustness.} By utilizing multiple factors in the procedure
of continuous authentication, it becomes difficult for adversaries to imitate
or crack all selected features in a single round of communication based on received signals and observations.

\section{Design of the Proposed ATMAS in SAGINs\label{sec:Design}}

\begin{figure*}
\centerline{\includegraphics[width=35pc]{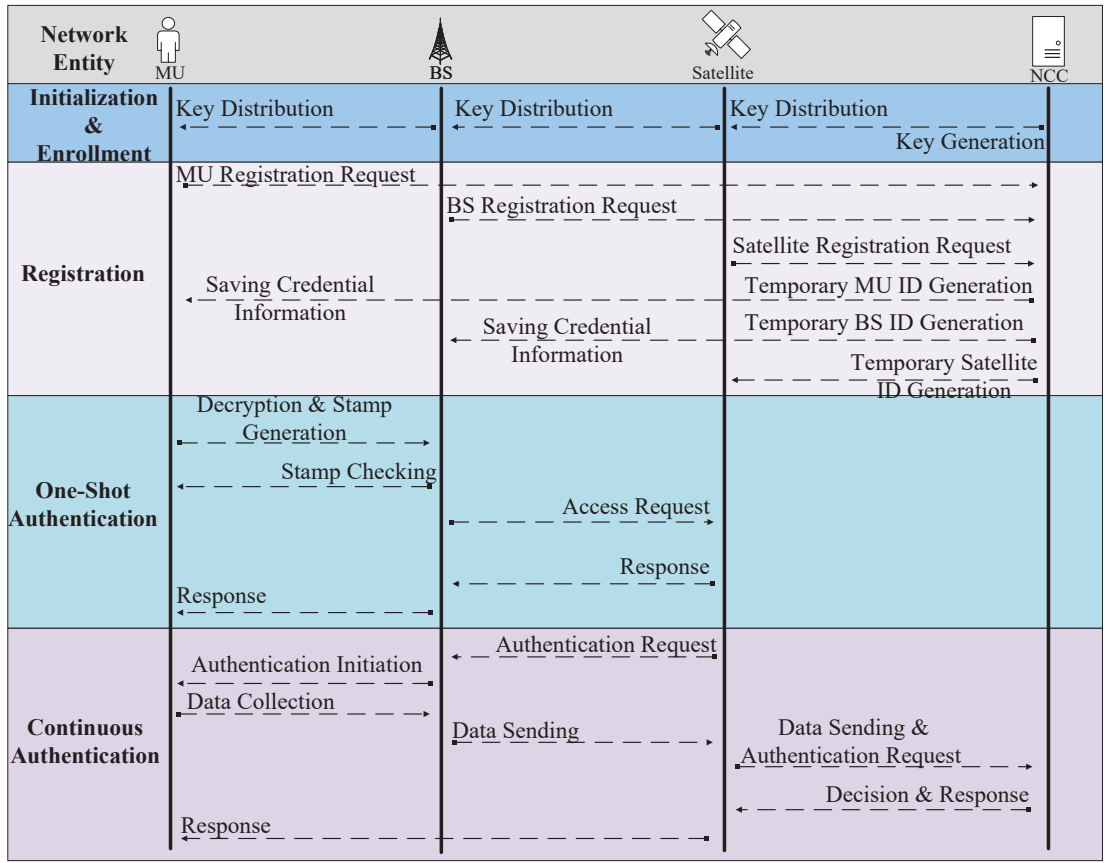}}
\caption{\label{fig:Procedure}Authentication Procedure of the ATMAS. The ATMAS consists of four procedures, i.e., initialization and enrollment, registration, one-shot authentication, and continuous authentication.}
\end{figure*}

Before introducing the detailed design of the proposed ATMAS, we would
like to provide an overview of the architecture of the SAGIN, as illustrated
in Fig. \ref{fig:Architecture} and Fig. \ref{fig:Phase}. The SAGIN system consists of three segments: the user/device segment, the ground/air/space access segment and the authentication segment. The user/device segment
is an assemblage of various end mobile users and IoT devices, and
it is assumed that users and devices have the ability to sense their
locations, traffic and biometrics. The ground/air/space access segment
consists of all kinds of access entities, such as terrestrial BSs, high-altitude platforms and satellites. The authentication
segment is responsible for one-shot authentication and continuous
authentication. In this article, the focus is on MEO/LEO satellites, which
have relatively shorter transmission delays compared to the GEO satellites.
This allows satellites to provide large-scale coverage, and beams
of BSs or other relays ensure high-precision user targeting \cite{Fang215G}.
The authentication procedure of the proposed ATMAS is shown in Fig. \ref{fig:Procedure},
in which four network entities are involved: MU,
BS, satellite, and NCC. In the following
subsections, the authentication procedure will be discussed in detail.

\subsection{Phase I: One-shot Authentication}
We divide authentication of Phase I into three procedures: initialization
and enrollment, registration and one-shot authentication.

\textbf{Initialization and Enrollment:} The authentication in Phase
I is cryptography-based. NCC uses Elgamal encryption technology \cite{ElGamal85A}
to generate a private key and a corresponding public key. The private
key is stored by the NCC locally, and the public key and its related key
parameters are published to other entities in SAGINs. Based on the
key parameters, other entities can generate their own private keys
and public keys, respectively.

\textbf{Registration:} Before participating in the authentication
procedure, entities including MUs, BSs and satellites must register with the NCC. The BS and the satellite have
a similar registration procedure and we take satellite registration
as an example. After receiving a registration request from a satellite,
the NCC generates both a unique identity and a temporary identity for the
satellite. Authentication parameters, such as satellite identities and
shared public keys, are stored locally on the satellite. Next, the
satellite generates a timestamp and sends it, along with authentication
parameters, to the NCC. Upon receiving this information, the NCC first verifies whether the transmission delay (i.e., the difference between the current time
and the sent timestamp) is less than the predefined delay threshold.
If the transmission delay does not meet the requirement, the NCC rejects
the registration request. Otherwise, the NCC checks where the authentication parameters sent by the satellite coincide
with parameters stored in the NCC's local database. If the parameters do not match, the satellite registration request is rejected. Otherwise, another timestamp is generated and
sent together with the authentication parameters back to the satellite. When
the satellite receives the message from the NCC, it first verifies
whether the transmission delay from the NCC to the satellite meets
the requirements. If so, the satellite also verifies the authentication parameters received from the NCC. The satellite registration procedure is considered validated if all checks pass.

For MU registration, we assume that MU and IoT devices have the ability
to collect raw biometric data. Firstly, the user selects an identity
and a high-entropy password, and utilizes a fuzzy extractor probabilistic
key generation algorithm to extract the key of biometric information.
The MU uses a hash function to generate authentication parameters
of the combination of identity, password and biometric features.
The authentication parameters are transmitted to the NCC. After receiving
the information, the NCC verifies whether the authentication parameters
are the same as those stored locally. If so, the authentication parameters
will be calculated by a hash function and sent to the MU. The MU authentication procedure is accomplished if the authentication parameters
are validated.

\textbf{One-Shot Authentication:} In this phase, when the MU requests
to access the SAGIN, the MU must provide the identity information,
a password, and biometric  information for a handheld device. Then, the
device generates a timestamp and obtains the MU's current location.
A hash function is utilized to calculate the key based on the combination
of the above information. The key is then sent to the satellite through
the BS. After receiving the key, the satellite first verifies whether
the transmission delay is within the threshold. If not, the MU access
request is denied. Otherwise, the satellite sequentially verifies the identity,
location, password, and biometric information. If the validation succeeds, authentication parameters and a timestamp are generated and transmitted to the MU through the BS. Similarly, the MU checks
the timestamp and authentication parameters. If matched, the MU
is marked as a legal user and allowed to access the SAGIN.

\subsection{Phase II: Continuous Authentication}
After the user gets access to the SAGIN, the NCC initiates
a continuous authentication request and a biometric/behavioral data
acquisition request to the MU. The collected data is then sent to
the NCC through the BS. In the MU registration phase, the NCC trains
an AI-based model and extracts features of the MU. In the continuous
authentication phase, the NCC compares the real-time extracted features
with the stored features from the registration phase. Based on the security level of the MU, the NCC determines the legality of the user, and the transmission terminates if it is an illegal user.

Moreover, due to the latency requirement of the authentication, it
is arduous to use high-complexity algorithms. In phase II, a binary
classification is introduced where both spoofing actions and spoofers are seen as spoofing attacks. For the sake of simplicity, we focus
on this scenario as a quintessential example to validate the proposed
authentication scheme.

\section{Security Analysis}
This section evaluates the security of the proposed authentication scheme and demonstrates its ability to prevent security threats. In our proposed ATMAS, we have implemented five security characteristics: mutual authentication, forward security, resistance to replay attack, resistance to man-in-the-middle attack, and data confidentiality and integrity. In addition, to demonstrate the effectiveness of our proposed authentication scheme, we compare it with existing counterparts in terms of security requirements and authentication features. By doing so, we can show how our scheme offers improved security and more robust authentication capabilities in comparison to existing solutions.

\begin{table*}
\centering
\caption{\label{tab:Comparison}Comparisons of Security Requirements and Authentication Features.}
\begin{tabular}{|c|c|c|c|c|}
\hline
 & Li et al. \cite{Li17A} & Deng et al. \cite{Deng21A} & Badhib et al. \cite{Badhib21A} & Our Scheme\tabularnewline
\hline
\hline
Authentication Phase & One & One & Two & Two\tabularnewline
\hline
Method & Crypto & Crypto & Crypto & Crypto, AI\tabularnewline
\hline
Dynamic & $\times$ & $\times$ & $\surd$ & $\surd$\tabularnewline
\hline
Mutual Authentication & $\times$ & $\times$ & $\surd$ & $\surd$\tabularnewline
\hline
Forward Security & $\surd$ & $\surd$ & $\surd$ & $\surd$\tabularnewline
\hline
Resistance to Replay Attack & $\surd$ & $\surd$ & $\surd$ & $\surd$\tabularnewline
\hline
Resistance to Man-in-the-Middle Attack & $\surd$ & $\surd$ & $\surd$ & $\surd$\tabularnewline
\hline
Data Confidentiality and Integrity & $\surd$ & $\surd$ & $\surd$ & $\surd$\tabularnewline
\hline
\end{tabular}
\end{table*}

\begin{itemize}
  \item \textbf{Mutual Authentication:} Mutual authentication implies that two participants can authenticate each other. In the key agreement phase described in one-shot authentication, the satellite can verify the legitimacy of the BS by checking whether the hash result is consistent with the credential information saved locally. Adversaries cannot get knowledge of the private key from public values due to the properties of hash function. Thus, only the legitimate BS who owns the private key can be authenticated by the satellite. Similarly, the BS authenticates the satellite based on the authentication messages from the satellite, which are encrypted by the satellite's private key. The secure mutual authentication between a MU and a BS can be guaranteed using a similar analysis. Therefore, our scheme could provide secure mutual authentication.
  \item \textbf{Forward Security:} Forward secrecy can ensure that session keys will not be compromised even if the long-term secrets used in the session key exchange are compromised. In our scheme, if an adversary learns the initial token and wants to derive the initial token used in the previous session, they would need to know the previously generated random numbers. However, the previously generated random numbers cannot be obtained from the previous eavesdropped messages as the hash is a one-way function. Therefore, our proposed scheme has the property of forward secrecy.
  \item \textbf{Resistance to Replay Attack:} Replay attack mainly refers to when a malicious adversary uses its message regeneration ability to generate and replay a message, thereby compromising protocol security. However, this attack can be prevented by successfully authenticating the message through checking the validity of its timestamp value. Additionally, because the hash function used in initial authentication is unidirectional, an adversary is unable to fake the message by modifying the timestamp value. Therefore, our proposed scheme is capable of resisting replay attack.
  \item \textbf{Resistance to Man-in-the-Middle Attack:} A man-in-the-middle attack means that an adversary intercepts and selectively modifies communicated data to masquerade as one entity involved in a communication session. In our scheme, it is impossible for the adversary in the middle to register with the role that already exists in the SAGIN. Besides, the secret keys of any entities in the system cannot be obtained. As a result, the adversary cannot modify or manipulate transmitting messages to invade the existing connection. Therefore, the scheme would not be exposed to man-in-the-middle attacks.
  \item \textbf{Data Confidentiality and Integrity:} Data confidentiality and integrity imply that a message receiver can ensure the message has not been tampered with during transmission. In our proposed scheme, the private information is encrypted with the public key of the target entity. An adversary cannot decrypt the information without the corresponding private key. Therefore, our proposed scheme achieves data integrity property.
\end{itemize}

Moreover, we also compare our proposed ATMAS with its counterparts against security requirements and authentication features in Table \ref{tab:Comparison}. From this table, the work in \cite{Li17A}-\cite{Badhib21A} adopts crypto-based authentication schemes, in which the schemes in \cite{Li17A} and \cite{Deng21A} have no authentication at the user device side. The work in \cite{Badhib21A} and our scheme use a two-phase authentication method, but the second phase in \cite{Badhib21A} is still crypto-based, which is less robust than ours.

\color{black}
\section{Case Study and Performance Evaluations\label{sec:Case}}
In this section, we first provide the factors/features used in our
analysis. Then, a comparison of the performance of classic AI-based
algorithms using selected factors is presented. Lastly, we discuss
the effect of factor selection strategy on authentication accuracy.

In our case study, we consider three kinds of services: conversational, streaming, and interactive. We assume that the BSs in the
SAGIN are fixed and uniformly distributed, and the coverage range of a BS is 20 $km$. The satellites in the SAGIN orbit at a height of 20000 $km$ and have a maximum beam scanning range of $\pm11.64{^\circ}$.
Table \ref{tab:Factors} provides the factors used in our proposed ATMAS,
including identities, trajectory, and communication attributes.

\begin{table}
\caption{\label{tab:Factors}Descriptions of Factors in the ATMAS.}

\begin{tabular}{|c|c|}
\hline
Items & Factors\tabularnewline
\hline
\hline
Factor 1 & Traffic Volume\tabularnewline
\hline
Factor 2 & Service Type\tabularnewline
\hline
Factor 3 & Uplink Rate\tabularnewline
\hline
Factor 4 & Sinuosity of the MU\tabularnewline
\hline
Factor 5 & Index of the BS\tabularnewline
\hline
Factor 6 & Distance between the BS and the MU\tabularnewline
\hline
Factor 7 & Position of the MU\tabularnewline
\hline
Factor 8 & Heading azimuth of the MU\tabularnewline
\hline
Factor 9 & Elevation between the BS and the Satellite\tabularnewline
\hline
\end{tabular}
\end{table}

\begin{figure}
\vspace*{-10pt}
\centerline{\includegraphics[width=17.5pc]{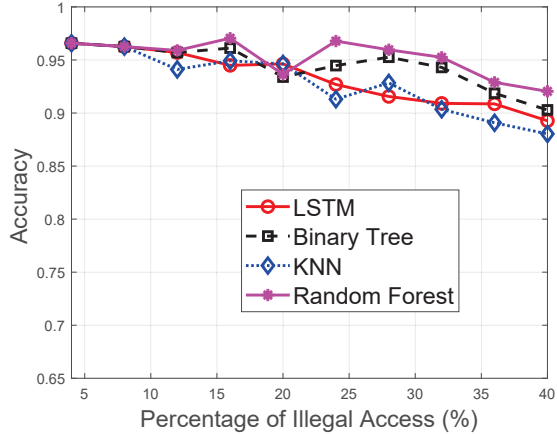}}
\caption{\label{fig:Accuracy}A Comparison of Authentication Accuracy with
Classic AI-based Algorithms.}
\end{figure}

\begin{figure}
\vspace*{-10pt}
\centerline{\includegraphics[width=17.5pc]{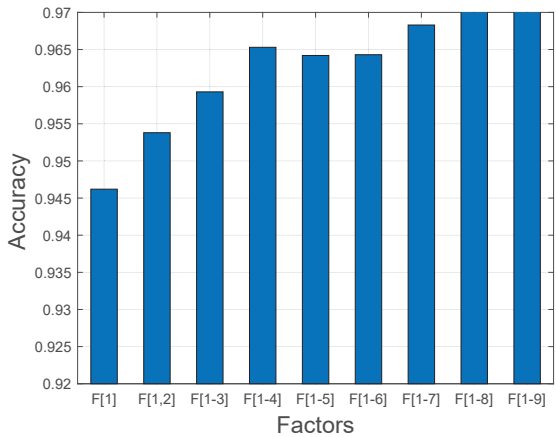}}
\caption{\label{fig:Number}Authentication Accuracy of the SAGIN Using a Different
Number of Factors.}
\end{figure}

\begin{figure}
\vspace*{-10pt}
\centerline{\includegraphics[width=17.5pc]{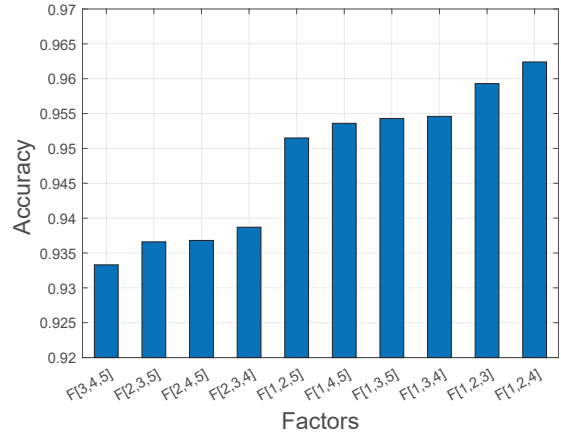}}
\caption{\label{fig:Random}Authentication Accuracy of the SAGIN Using Random
Selected Factors.}
\end{figure}

We compare the authentication accuracy of different AI-based algorithms
with respect to the percentage of illegal access. The authentication accuracy
is defined as $ACC=\left(TP+TN\right)/\left(TP+FP+TN+FN\right)$, where $TP$, $FP$, $TN$, and $FN$ denote true positives, false positives, true negative, and false negative, respectively.
As shown in Fig. \ref{fig:Accuracy}, with an increase in the percentage
of illegal access, the authentication accuracy decreases slightly by less than 8\%. Besides, based on the random forest
algorithm, the authentication accuracy can achieve over 92\%.
By using the random forest algorithm, the proposed authentication
scheme is robust even if the percentage of illegal access is high.

Fig. \ref{fig:Number} illustrates the authentication accuracy using
different numbers of factors. F{[}1{]} means that we only choose
Factor 1 and F{[}1-9{]} means that all factors are selected. The results
show that as the number of selected factors increases, the
authentication accuracy also increases. However, the increase rate decreases, such that the authentication accuracy using 8 factors
is almost the same as that using 9 factors. This indicates that not
every factor has the same contribution to the increase in authentication
accuracy. Therefore, choosing the proper combination of factors makes lightweight and effective authentication possible. In Fig. \ref{fig:Random},
we present the authentication accuracy of a combination of 3 factors.
The results show that the combination with Factor 1, Factor
2 and Factor 4 has the highest authentication accuracy. Besides, combinations
with Factor 1 can achieve an average accuracy of more than 95\%, which
indicates that in our simulation scenario, Factor 1 contributes the
most to the ATMAS.

\section{CONCLUSION\label{sec:Con}}
In this article, an ATMAS is proposed in which spatial-temporal features
are taken into account, including traffic volume and geographic location. An AI-based continuous
authentication, referred to as `Phase II', is performed to intelligently
grant or deny access reliably by capturing user profiles and traits following a conventional cryptographic authentication named `Phase
I'. Security analysis is conducted to evaluate the robustness and validate the security of the proposed
Scheme. Performance evaluation shows that the mean authentication accuracy
achieves over $92\%$ by ATMAS. Additionally, we analyze
the importance of factors used in our case study, shedding light
on lightweight and efficient authentication mechanism design. However,
there are still several research opportunities for future study.

\textbf{Authentication Centers Deployment.} In our proposed ATMAS, authentication is completed by the NCC on the ground. However, this increases transmission
delays since all authentication information must be transmitted to the
NCC. To address this, choosing MEO satellites as partial authentication centers is
a promising solution. However, communication and computation resources
on the satellite are limited, which requires optimization of the allocation
of authentication tasks between MEO satellites and the NCC.

\textbf{Efficient AI-based Authentication.} AI-based authentication
algorithms have intensive computation and communication costs and
require a large amount of training data as well as a complicated feature-extraction
process \cite{Xiao18IoT}. Besides, low computation and communication overhead algorithms should be investigated to the deployment of authentication centers on satellites.

\textbf{Blockchain-based Intelligent Authentication.} Blockchain is
a distributed ledger technology with characteristics of decentralization,
security, interoperation, and trust establishment \cite{Adhikari22Security}.
As the SAGIN becomes increasingly complex, the system may suffer
from false authentications leading to potential privacy leakage
and security risks. Blockchain-based techniques can be utilized to
track past security breaches and provide the necessary log analysis.

\section{ACKNOWLEDGMENTS}

This work was supported in part by the National Key Research and Development Program of China under Grant 2020YFB1804800 and Grant 2021YFB2900200, in part by the Key Research and Development Program of Zhejiang Province under Grant 2021C01197, in part by the National Natural Science Foundation of China under Grant 62101509, in part by the Natural Science Foundation of Zhejiang Province under Grant
LQ22F010018. The corresponding authors of this article are Yongdong Zhu and Zipeng Li.

\begin{IEEEbiography}{Bin Yang}{\,}received the Ph.D. degrees in information and communication engineering from the Huazhong University of Science and Technology, Wuhan, China, in 2018. His research interests include vehicular communication networks, and reconfigurable intelligent surface assisted communications. Contact him at binyang@zhejianglab.com.
\end{IEEEbiography}

\begin{IEEEbiography}{Shanyun Liu}{\,} received the B.S., M.S., and Ph.D. degrees from the Department of Electronic Engineering, Tsinghua University, Beijing, China, in 2014, 2016 and 2020, respectively. His current research interests include integrated terrestrial-satellite communications, graph learning and information theory. Contact him at liusy@zhejianglab.com.
\end{IEEEbiography}

\begin{IEEEbiography}{Tao Xu} {\,} received the M.Sc. and Ph.D. degree in Aberystwyth University, Abersytwyth, UK, in 2015 and 2021, respectively. He is currently a postdoctoral researcher in Zhejiang Lab, Hangzhou, China. His current research interests include evolutionary computation, constrained optimization, machine learning, and graph computing.  Contact him at xut@zhejianglab.com.
\end{IEEEbiography}

\begin{IEEEbiography}{Chuyu Li} {\,} received her B.E. and Ph.D. degrees from the University of Wollongong, Australia, in 2017 and 2020, respectively. She joined Zhejiang Lab, China in 2021 as a postdoctoral researcher. Her current research interests include UAV-aided sensor networks, radar-radio systems and RF-energy harvesting networks. Contact her at lcy@zhejianglab.com.
\end{IEEEbiography}

\begin{IEEEbiography}{Yongdong Zhu} {\,} received his Ph.D. degree from the University of Essex, Colchester, United Kingdom, in 2007. He is currently a professor with the Zhejiang Lab. His current research interests include next generation mobile communication, mobile edge network architecture, vehicular communication networks, and Internet of Things. Contact him at zhuyd@zhejianglab.com.
\end{IEEEbiography}

\begin{IEEEbiography}{Zipeng Li} {\,} received his Ph.D. degrees in information and communication engineering from the Huazhong University of Science and Technology, Wuhan, China, in 2021. He is currently a Lecturer with Nanchang University, Nanchang, China. His research interests include vehicular networks, artificial intelligence, and autonomous vehicles. Contact him at zipengli@ncu.edu.cn.
\end{IEEEbiography}

\begin{IEEEbiography}{Zhifeng Zhao} {\,} received his Ph.D. degree in communication and information system from the PLA University of Science and Technology, Nanjing, China, in 2002. Currently, he is with Zhejiang Lab. His research area includes software defined network, wireless network in 6G, computing network, and collective intelligence. Contact him at zhaozf@zhejianglab.com.
\end{IEEEbiography}

\end{document}